\documentclass[10pt,twocolumn,twoside]{IEEEtran}
\usepackage{amssymb,amsmath,graphicx,bm}
\usepackage{pbox}
\usepackage{ragged2e}
\usepackage{hyperref}
\usepackage{url}
\usepackage[table]{xcolor}
\usepackage{float}
\usepackage{placeins}
\usepackage{soul}
\usepackage{subfigure}
\usepackage{dblfloatfix} 
\usepackage[ruled, lined, linesnumbered, commentsnumbered, longend]{algorithm2e}
\begin{document}

\title{Vocal Tract Length Perturbation for Text-Dependent Speaker Verification with Autoregressive Prediction Coding}
\author{Achintya kr. Sarkar, Zheng-Hua Tan, \emph{Senior Member, IEEE}
\thanks{A. K. Sarkar is with Indian Institute of Information Technology, Sri City, India (E-mail: sarkar.achintya@gmail.com). Z.-H. Tan is with Department of Electronic Systems, Aalborg University, Denmark (E-mails: zt@es.aau.dk).}}

\makeatletter
\def\ps@IEEEtitlepagestyle{%
  \def\@oddfoot{\mycopyrightnotice}%
  \def\@evenfoot{}%
}
\def\mycopyrightnotice{%
  {\footnotesize Copyright $\copyright$ 2021 IEEE. This article is the \emph{accepted} version of IEEE Signal Processing Letters.
DOI:10.1109/LSP.2021.3055180. \hfill}
  \gdef\mycopyrightnotice{}
}

\maketitle

\begin{abstract}
In this letter, we propose a vocal tract length (VTL) perturbation method for  text-dependent speaker verification (TD-SV), in which a set of TD-SV systems are trained, one for each VTL factor, and score-level fusion is applied to make a final decision. Next, we explore the bottleneck (BN) feature extracted by training deep neural networks with a self-supervised learning objective, autoregressive predictive coding (APC), for TD-SV and comapre it with the well-studied speaker-discriminant BN feature. The proposed VTL method is then applied to APC and speaker-discriminant BN features. In the end, we  combine  the VTL perturbation systems trained on MFCC and the two BN features in the score domain.
Experiments are performed on the RedDots challenge 2016 database of TD-SV using short utterances with Gaussian  mixture model-universal background model and i-vector techniques. Results show the proposed methods significantly outperform the baselines. 
\end{abstract}



\IEEEpeerreviewmaketitle

\section{Introduction}
Speaker verification (SV) is the task of verifying a person using his/her speech signal, which can be either text-independent (TI) or text-dependent (TD). In a TI-SV system, speakers are free to speak any text content during the system enrollment and test phases. On the contrary, TD-SV constrains a speaker to speak a particular pass-phrase from a predefined list in both enrollment and test phases. As TD-SV maintains the matched phonetic contents between training and test phases, it is superior to TI-SV for SV with short utterances and ideal for real-world applications. 

An important factor that characterises speech signals is vocal tract length (VTL), which varies from person to person and is of importance to deal with in speech systems including speaker recognition and automatic speech recognition (ASR). For example, vocal tract length normalization (VTLN) \cite{7078563,umesh_sadana,akhil-interspeech2008} is widely used in ASR to reduce the effect of speaker variability due to the difference in VTL among speakers. In \cite{Jaitly_vocaltract}, VTL warped features are used for ASR as a data augmentation technique. 

 
The use of VTL information in speaker recognition is rather limited with few applications in \emph{text-independent} context only, e.g., speaker identification \cite{sarkar2009} and verification \cite{achintya-odyssey2010}.
In \cite{sarkar2009}, VTL warped features are used for building a Gaussian mixture model-universal background model (GMM-UBM) called VTLN-UBM for speaker identification. Then speaker models are derived from the GMM-UBM using the enrolment data of each speaker. During test, feature vectors of a test utterance is scored against the models of registered speakers to identify the best matching speaker. Only training data are warped, but enrollment and test data are the original cepstral features without warping. In \cite{achintya-odyssey2010, Zhang-odyssey2010}, different GMM-UBM models are trained based on the VTL 
factors of speakers in training data; each model corresponds to one VTL factor value. During enrolment, speaker models are derived from the VTL factor-specific GMM-UBM that matches their VTL factor values by using enrolment data. During test, the feature vectors of the test utterance are scored against the claimant speaker model and the corresponding GMM-UBM. All training, enrollment and test data are unwarped.

In summary, VTL warped features have been used for TI speaker identification and the VTL concept has been exploited for TI-SV. To the best of our knowledge, no prior work exploits VTL warped features for SV. Furthermore, no prior work applies the VTL concept for TD-SV. As VTL is an important factor characterising speakers, \emph{we propose a VTL perturbation method for TD-SV.} Specifically, the proposed method trains a set of TD-SV systems, one for each VTL factor, and decision level fusion is applied to combine the systems to make final SV decision. Note that the VTLN-UBM method in \cite{sarkar2009} is not suitable for SV as VTLN-UBM is trained with warped features whereas targets are derived with unwarped features during adaptation. Therefore a test unwarped feature will mismatch with the VTLN-UBM.  Paper \cite{achintya-odyssey2010} extends the concept of gender dependent SV further by grouping the speaker based on their VTL values, which can be used in conjunction with the proposed method. 

Recently, self-supervised learning has attracted significant attention in the community. A notable method of such is based on autoregressive prediction coding (APC), and it has shown to be able to learn powerful speech representations, which perform extremely well for several downstream tasks including  ASR and speaker identification \cite{chung2020generative, Chung2019}. In APC, a deep neural network (DNN) encoder is trained with the objective function to predict future frames. The output from a selected hidden layer is considered as the APC feature. It is shown in \cite{chung2020generative}  that the APC feature significantly outperforms the widely-used cepstral feature. However, \emph{the performance of APC on TD-SV has not been established yet}. Besides, we compare it to a bottleneck (BN) feature trained in a supervised fashion. In the literature, there are a number of BN features such as those trained to  discriminate speakers \cite{Yuan2015},  senones \cite{McLaren2015}, a combination of them \cite{Yuan2015}, or tri-phone state \cite{Yuan2015}, time-contrastive-learning (TCL) and phone discriminate \cite{DBLP:journals/taslp/SarkarTTSG19}. In this work, we choose the BN feature discriminating speakers (Spk-BN) for comparison.

To investigate joint effect of BN features and VTL perturbation on TD-SV, we apply the proposed VTL method to the BN features including both APC-BN and Spk-BN. In the end, we fuse the systems based on MFCC, APC-BN and Spk-BN in the score domain. 

We conducted experiments on the RedDots dataset of TD-SV using short utterances \cite{RedDots} and show encouraging results under both GMM-UBM and i-vector modeling frameworks. The choice of these modelling methods is based on the comparison in \cite{UIAI20}, which shows that GMM-UBM remains a better choice for TD-SV using short utterances than other methods. Although x-vector \cite{DBLP:conf/icassp/SnyderGSPK18} has shown promising results for TI-SV, it is not successful so far for TD-SV possibly due to limited training data~\cite{zeinali2019short}.


\section{Features and TD-SV modeling methods}
\label{sec:tdSV}
\subsection{Features}
\subsubsection{Spk-BN feature} During training, Mel-frequency cepstral coefficients (MFCCs) are fed to a DNN to discriminate speakers with coss-entropy objective function. 
When using it, the output of a particular hidden layer of the DNN for a given speech segment at frame-level is projected onto the low dimensional space using principal component analysis (PCA) to get Spk-BN feature {\cite{DBLP:journals/taslp/SarkarTTSG19}}. 

\subsubsection{APC-BN feature}
 During training, MFCC feature vectors are fed to a DNN encoder with a objective function to minimize the $L1$ loss between the predicted sequence $(\bm{y}_1 , \bm{y}_2, \ldots, \bm{y}_N)$ and the target sequence $(\bm{t}_1, \bm{t}_2, \ldots \bm{t}_N)$ by right-shifting the input sequence $(\bm{x}_1, \bm{x}_2, \ldots, \bm{x}_N)$ of $n$ time steps:
 
 \begin{equation}
     \sum_{i=1}^{N-n} |\bm{t}_i -\bm{y}_i|, \quad  \bm{t}_i = \bm{x}_{i+n}.
     \label{eq:apc}
 \end{equation} 
 When using it, the output from a particular hidden layer of the DNN for a given utterance is extracted at frame-level to get the APC feature of 512 dimensions, which is then used for text-independent speaker identification \cite{chung2020generative} and and verification \cite{Chung2019}. 
 In this work, the 512-dimensional APC feature is projected onto the low dimensional space using PCA to get APC-BN feature of the same dimension as Spk-BN. 

\subsection{Speaker verification methods}
\label{sec:GMM_tech}
\subsubsection{GMM-UBM technique} A GMM-UBM  \cite{reynold00} is trained using data from many non-target speakers. Speakers are then represented by their models derived from the GMM-UBM using the enrollment data of the particular speaker with maximum a posteriori (MAP) adaptation. During test, the feature vectors of a test utterance $\mathbf{X}=\left\{\mathbf{x}_1,\mathbf{x}_2,\ldots, \mathbf{x}_N \right\}$ are scored against the claimant model $\lambda_r$ (obtained in the enrollment phase) and GMM-UBM $\lambda_{ubm}$, respectively. Finally, a log-likelihood ratio (LLR) value $\Lambda(\bm{X})$ is calculated using scores between the claimant and GMM-UBM models. The LLR value is used to accept or reject the claimant.
\begin{eqnarray}
\Lambda(\mathbf{X}) = \frac{1}{N}\big[\log p(\mathbf{X}|\lambda_r)- \log p(\mathbf{X}|\lambda_{ubm}) \big] \label{eq:llr}
\end{eqnarray}



\subsubsection{i-vector} In this technique \cite{Deka_ieee2011}, a speech signal is represented by a vector called \emph{i-vector}. An i-vector is obtained by decomposing the speaker and channel dependent GMM super-vector $\bm{M}$ as 
\begin{equation}
 \bm{M} = \bm{m} + \bm{Tw} 
\end{equation}
where  $\bm{m}$ denotes the speaker-independent GMM super-vector and {$\bm{w}$ is called an i-vector}. $\bm{T}$ is the total variability space  (T-space) on  a subspace of $\bm{m}$, where speaker and channel information is assume to be dense.
 During the training phase, each target is represented by an average i-vector computed over his/her training utterance-wise (or session-wise) i-vectors. During test, the i-vector of the test utterance is scored against the claimant specific i-vector  (obtained during enrolment) with probabilistic linear discriminant analysis (PLDA) \cite{SenoussaouiInterspch2011}.



\section{Proposed VTL perturbation for TD-SV}
\label{sec:prop_methd}
\subsection{Concept of VTL}
In this concept \cite{Lee-Rose98}, the spectra for a spoken content between the two speakers (A and B) are matched  by scaling the frequency axis of the signal spectra,
\begin{eqnarray}
        S_A(f) = S_B(\alpha f)
        \end{eqnarray}
        
where, the $\alpha$ is called VTL  (or VTL warp) factor.
\begin{figure}[!b]
\centering{\includegraphics[height=6.0cm,width=7.2cm]{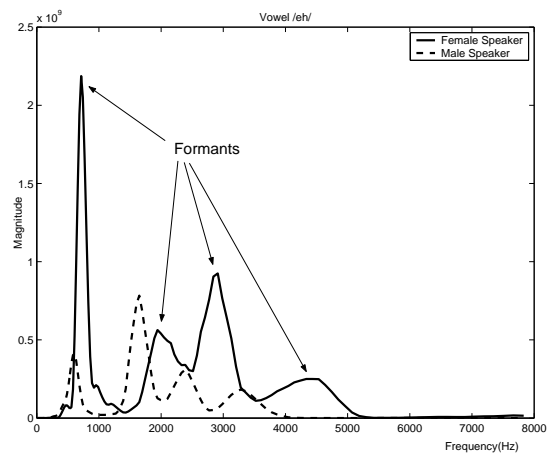}}
\caption{\it The spectra of vowel /eh/ for male and female speakers \cite{achintya-odyssey2010}.}
\label{fig:vtln_warp}
\end{figure}
Fig.~\ref{fig:vtln_warp} shows the smoothed spectra of vowel /eh/ uttered by a male and a female speaker.
To match the formants of the male speaker to those of the female speaker (as reference), it is required to scale the frequency axis, i.e., to expand the spectra of the male speaker.
Based on the psychological structure of male and female speakers, $\alpha$  generally lies in the range $[0.80, 1.20]$ with steps of $0.02$. To maintain the bandwidth of original signal ($f$) with warped frequency spectra ($f_{w}$), the warped frequency is expressed as a piece-wise linear \cite{Lee-Rose98},
\begin{eqnarray}
f_{w} = (\alpha f) = 
\begin{cases}
    \alpha f,      &  0 \leq f \leq f_0\\
    \frac{f_{max} -\alpha f_0}{f_{max}-f_0}(f-f_0) + \alpha f_0,  & \quad f_0 \leq f \leq f_{max}
  \end{cases}
\end{eqnarray}

\noindent where $f_{max}$ and $f_0$ denote the signal maximum bandwidth and empirical upper boundary frequency up to $\alpha$ linearly warped, respectively.

\subsection{Proposed method}
Conventional TD-SV systems use MFCC features with $\alpha=1.0$. In the proposed method, MFCC features are extracted for different values of VTL factors from $0.80$ to $1.20$ with increment value of $0.02$  by warping the frequency axis of spectra for a given utterance as per \cite{Lee-Rose98}. This gives $21$ sets of cepstral features (incl. $\alpha=1.0$) for a given speech utterance  and the sets of features are used for the development of TD-SV systems, separately, i.e., for training DNN, GMM-UBM, T-space and PLDA, leading to one system per VTL factor. 

When APC-BN and Spk-BN features are used, DNN models for BN feature extraction are trained using cepstral features warped for each VTL factor.
Afterward, the cepstral feature of an utterance for a particular VTL factor is fed to the corresponding DNN for Spk-BN and APC-BN. It can be expressed as
\begin{eqnarray}
\bm{x}_{\alpha}^{BN} & = & f_{DNN_{\alpha}}(\bm{x_{\alpha}}), \quad \alpha \epsilon \;[0.80, 1.20]
\end{eqnarray}

\noindent {where $\bm{x}^{BN}_{\alpha}$  denotes the Spk-BN or APC-BN  feature extracted from  MFCC with VTL factor $\alpha$ for a given speech utterance $x$ using the DNN corresponding to VTL factor $\alpha$. } The extracted  features are   used for building the TD-SV. It gives $21$ TD-SV systems for Spk-BN and APC-BN, respectively. Finally, scores of the different VTL feature based TD-SV systems are fused with equal importance and used for decision making,
\begin{eqnarray}
        Score_{fusion} = \frac{1}{N}\sum_{i=1}^{N} Score_i   \label{eq:fusion}
\end{eqnarray}

This is compared to systems use MFCC, Spk-BN and APC-BN features for  ($\alpha=1.0$). 

\begin{figure*}[t]
{\subfigure[\it]{\includegraphics[width=4.8cm,height=2.8cm]{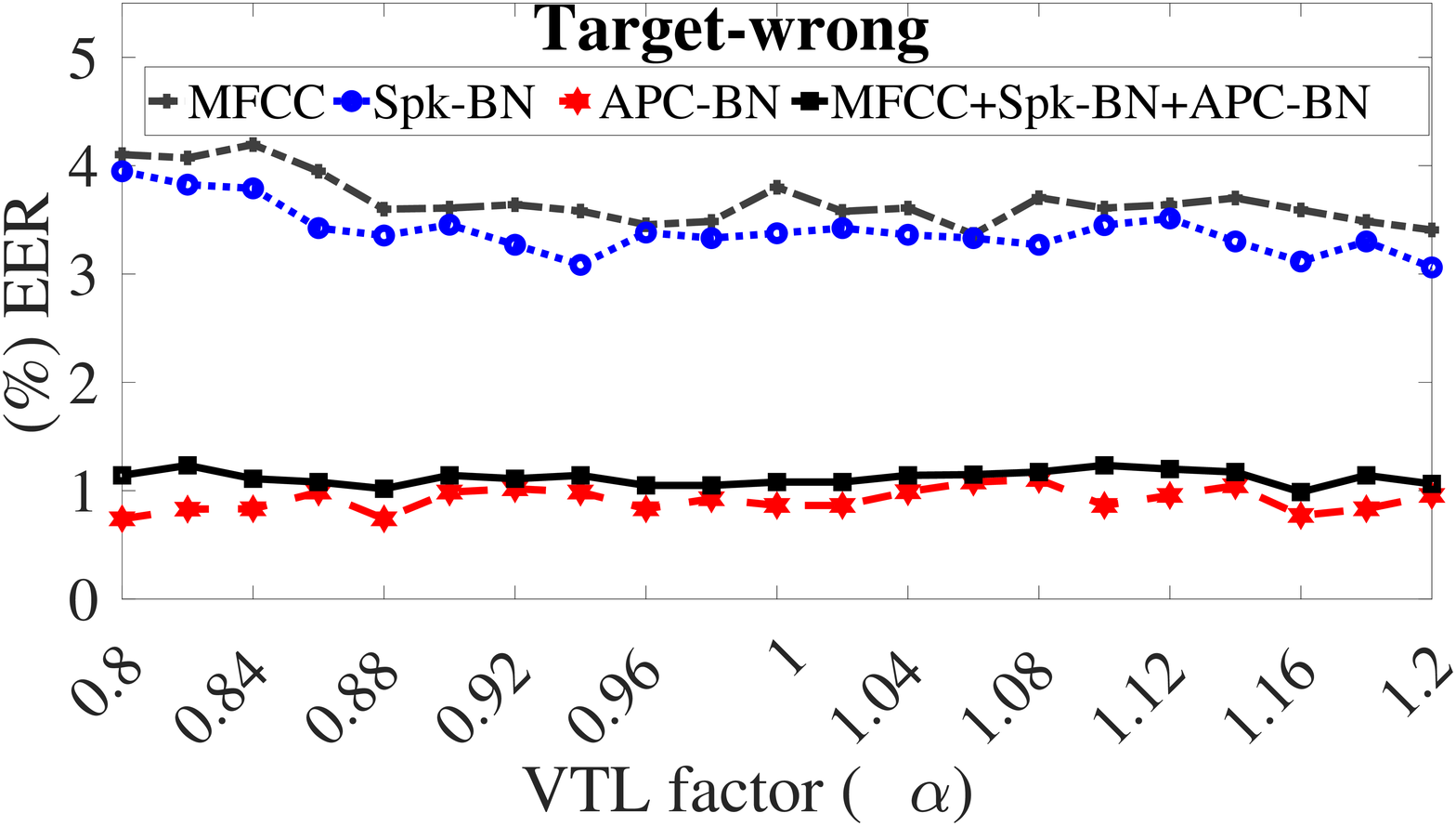}
\label{fig:46_anch}}}
{\hspace*{-0.60cm}\subfigure[\it]{\includegraphics[width=4.8cm,height=2.8cm]{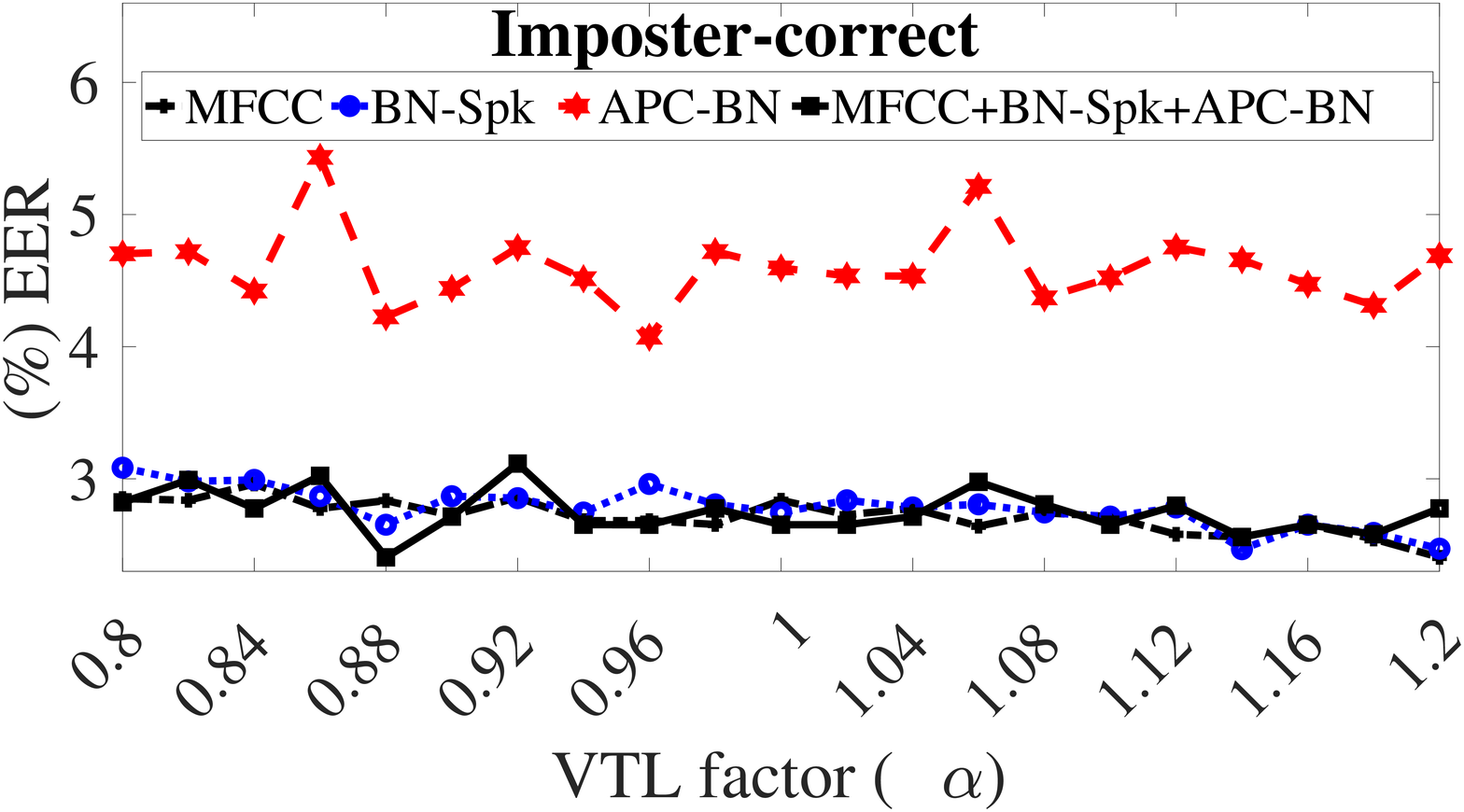}
\label{fig:146anch}}}
{\hspace*{-0.6cm}\subfigure[\it]{\includegraphics[width=4.8cm,height=2.9cm]{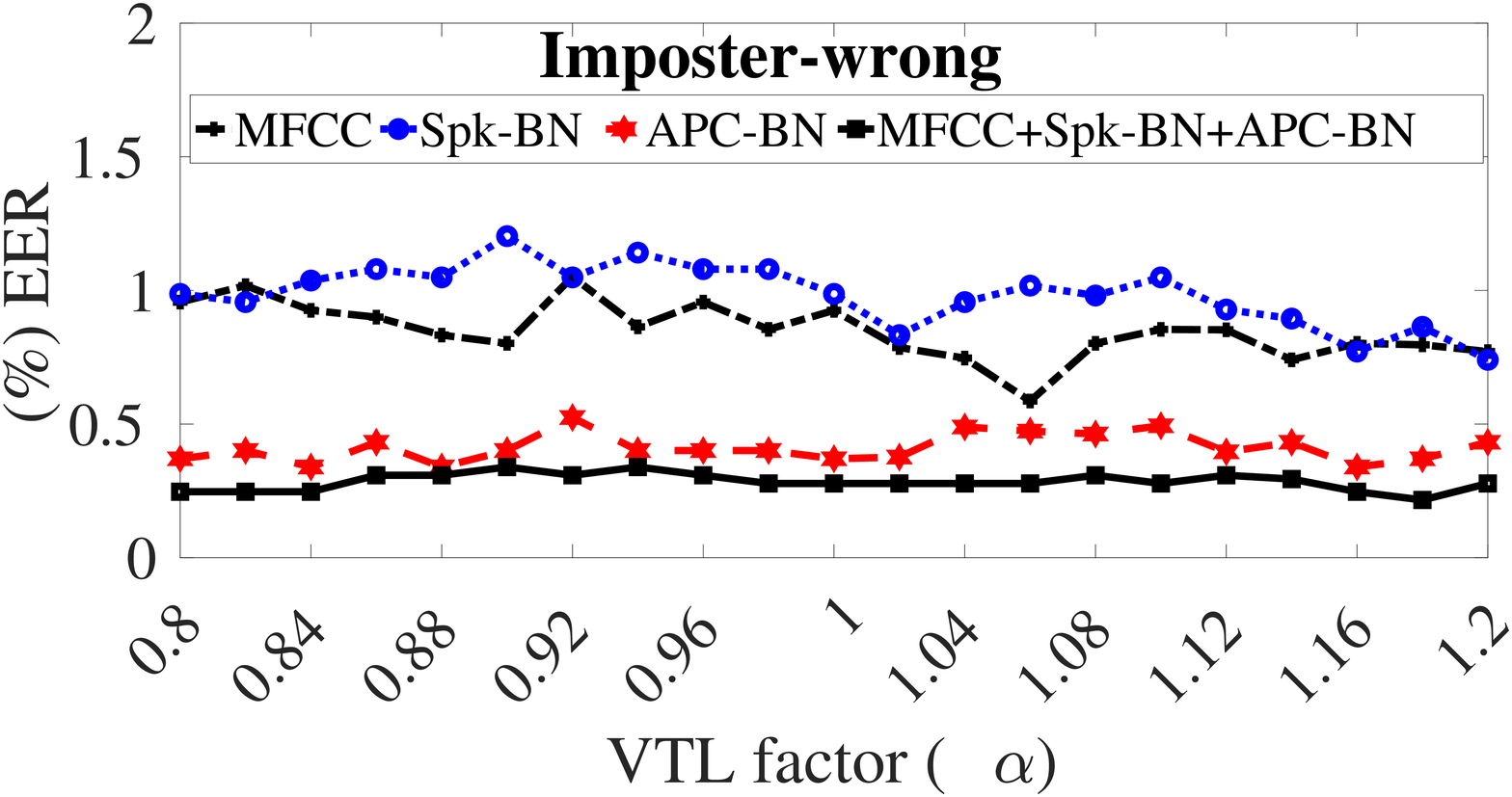}
\label{fig:246anch}}}
{\hspace*{-0.6cm}\subfigure[\it]{\includegraphics[width=4.8cm,height=2.8cm]{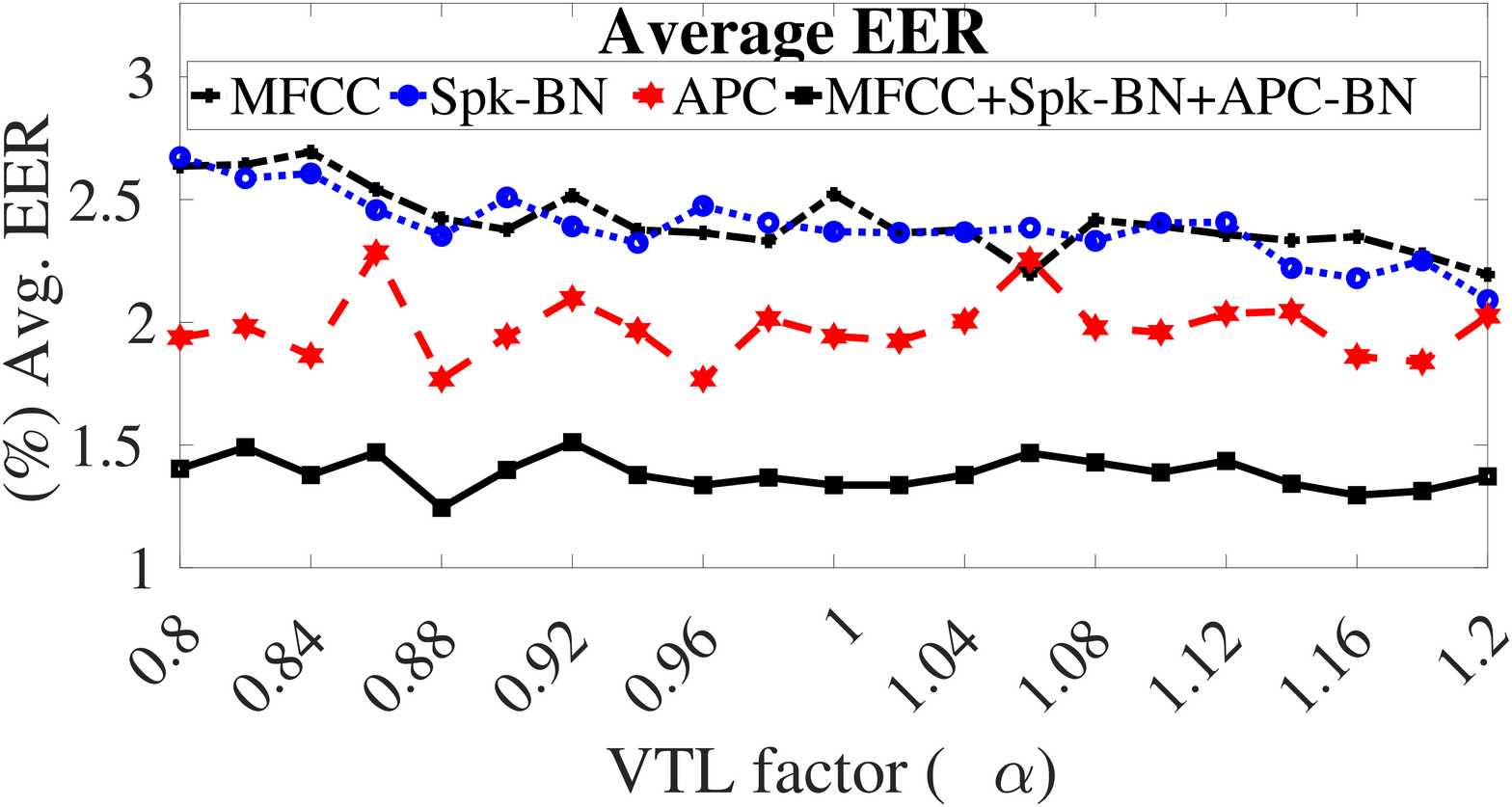}
\label{fig:346anch}}}
\caption{\it Comparison of TD-SV performance between MFCC, Spk-BN and APC-BN features and their score-level fusion across different values of $\alpha$ and different trials on the RedDots database with the GMM-UBM technique: {\bf (a)} Target-wrong {\bf(b)} Imposter-wrong {\bf(c)} Imposter-wrong  and {\bf(d)} Average EER over (a), (b) and (c).}  
\label{fig:sv_eer_alpha}
\vspace*{-0.3cm}
\end{figure*}

\section{Experimental setup}
\label{sec:expsetup}
Experiments for TD-SV are conducted on the \emph{m-part-01} task (for male speakers) of the RedDots database as per protocol  \cite{RedDots}. There are $320$ target models to train, each having three utterances/speech samples. Each utterance is approximately $2$ seconds duration on average. There are four types of trials for system evaluation shown in Table \ref{table:no_trial}. More details can be found in \cite{RedDots}.\\
\begin{itemize}
\item\emph{Genuine trial} is when a target speaker claims the identity during test by speaking a pass-phrase included in the enrollment. 
\item\emph{Target-wrong} is  when a target speaker claims the identity during test by speaking a pass-phrase that is not included in the enrollment.
 \item\emph{Imposter-correct} is when an imposter falsely claims as a genuine speaker with pronouncing a pass-phrase included in the enrollment.
\item\emph{Imposter-wrong} is when an imposter falsely claims as a genuine speaker with pronouncing a pass-phrase that is not included in the enrollment.
\end{itemize}

The RSR2015 database \cite{RSR2015} is used to train the total variability space and PLDA in the i-vector system and to train DNNs for Spk-BN and APC-BN feature extraction. The database consists of $157$ male and $143$ female speakers and of $72764$ utterances over $27$ pass-phrases after excluding the pass-phrases common with the RedDots database.

MFCC feature vectors of $57$ dimension (i.e., $19$ static and their $ \Delta, \Delta\Delta$) with  RASTA filtering \cite{Hermanksy94} are extracted from speech signals using a $25 ms$ hamming window and a $10ms$ frames shift for different values of $\alpha$. The  upper boundary frequency required in VTL \cite{Lee-Rose98} is set $85\%$ of highest frequency in the speech signal. HTK toolkit \cite{htkbook} is used for feature extraction. An open-source robust voice activity detection (rVAD)\footnote{\url{https://github.com/zhenghuatan/rVAD}} \cite{TanSD20} is applied to discard the less energized frames. The selected frames are then normalized to fit zero mean and unit variance at the utterance level.

\begin{table}[t]
\begin{center}
\caption{Number of trials in evaluation set.}
\vspace*{-0.1cm}
\begin{tabular}{|l|l|l|l|}\cline{1-4}
\# of  & \multicolumn{3}{c|}{\# of trials in Non-target type} \\ 
  Genuine            & Target  & Imposter & Imposter \\
  trials             &-wrong   &  -correct &  -wrong    \\   \hline
       3242       & 29178       & 120086            & 1080774  \\ \hline
\end{tabular}
\label{table:no_trial}
\end{center}
\end{table}  

A GMM-UBM with $512$ mixtures having diagonal co-variance matrices is trained using $6300$ speech files from the TIMIT database \cite{Timit} consisting of $438$ males and $192$ females.
This data set is also used for training low dimensional PCA space in BN feature extraction. In MAP adaptation, $3$ iterations with the value of relevance factor $10$, are considered.
In PLDA, the utterances of the same pass-phrase from a particular speaker are treated as an individual speaker. It gives $8100$ classes (4239 males and 3861 females) in PLDA. Speaker and channel factors are kept full  in PLDA, i.e. equal to the dimension of i-vector ($400$). Before PLDA, i-vectors are  normalized with spherical normalization \cite{Pierre-interspeech2012}.  

For Spk-BN, TensorFlow \cite{tensorflow2015-whitepaper} is used for training the DNNs {with the following settings: variable batch sizes from $256$ to $1024$,  variable learning rates from $0.8$ to $0.08$, and $30$ training epochs}. Seven layers of feed-forward networks with a sigmoid activation function are used. The input layer considers the context window of $11$ frames (i.e. $5$ frames left, current frame, $5$ frames right). Each hidden layer consists of $1024$ neurons. The number of nodes at output is equal to the number of speakers, i.e., $300$ in DNN training data. In Spk-BN, the frame-level output from the fourth hidden layer of DNNs is projected onto $57$ dimensional space to align with the dimension of the MFCCs feature for a fair comparison as per \cite{DBLP:journals/taslp/SarkarTTSG19}. 

Deep APC features are extracted as per \cite{chung2020generative}. DNN encoder consists of $3$ hidden layers GRU with ReLU activation function, the output layer is linear, the batch size is $32$, and the learning rate is $0.001$, and $n=5$ as in Eq \ref{eq:apc} for APC (which gives the best performance in \cite{chung2020generative}). Output from the last hidden layer of DNN at frame-level is projected onto $57$ dimensional space  to get APC-BN feature.  
System performance is measured in terms of equal error rate (EER) and minimum detection cost function (minDCF) as per the 2008 SRE \cite{SRE08}.

\begin{table}[!b]
\caption{\it TD-SV performance of the  proposed  VTL  perturbation  method  and  its  combination  with  APC-BN  and  Spk-BN features on the RedDots database in terms of \%EER/(MinDCF$\times$ 100) with GMM-UBM.} 
\setlength\tabcolsep{1.7pt}
\begin{center}
\scriptsize{
\begin{tabular}{|llll|l|}\cline{1-5}
System & \multicolumn{3}{c|}{Non-target type} & Avg. \\
       & Target-      & Imposter & Imposter- & EER/\\ 
      &  wrong        & correct  &  wrong    & MinDCF\\ \hline
{\bf Baseline:}  & & & &\\
MFCC  ($\alpha$=$1.0$)  & 3.80/1.38& 2.83/1.19&0.92/0.28  & 2.52/0.95 \\
Spk-BN ($\alpha$=$1.0$)    & 3.37/1.34 & 2.74/1.27 &0.98/0.28  & 2.36/0.96  \\ 
APC-BN  ($\alpha$=$1.0$)  & 0.86/0.29 &	4.59/2.03 &	0.37/0.09 &	 1.94/0.81 \\

&    & & & \\  
{\bf Proposed:} & & & & \\ 
MFCC (all $\alpha$)  & 3.06/1.23& 2.19/1.00&0.52/0.14 & 1.92/0.79 \\
Spk-BN (all $\alpha$)     &2.52/1.03 & {\bf 2.03/0.89} &0.49/0.13  & 1.68/0.68\\ 
APC-BN (all $\alpha$)    & {\bf 0.64/0.19}  &   3.80/1.71 &  0.27/0.05        &  1.57/0.65\\
& & & & \\ 
{\bf Score fusion:} & & & & \\ 
MFCC + Spk-BN ($\alpha$=$1.0$) & 3.26/1.22 & 2.51/1.09&0.77/0.19 & 2.18/0.84 \\
MFCC + APC-BN ($\alpha$=$1.0$) & 0.95/0.35         &  3.08/1.43             & 0.30/0.07   & 1.44/0.62\\
MFCC + [Spk + APC]-BN ($\alpha$=$1.0$) & 1.07/0.38         &2.65/1.26  & 0.27/0.07   & 1.33/0.57 \\
& & & & \\ 

MFCC + Spk-BN(all $\alpha$)   & 2.74/1.10 & 1.91/0.91     & 0.46/0.11 & 1.70/0.71 \\
MFCC + APC-BN (all $\alpha$)  & 0.85/0.25          & 2.83/1.30    &{\bf 0.24/0.04 }   & 1.31/0.53  \\
MFCC + [Spk + APC]-BN (all $\alpha$)   &  0.92/0.30        &   2.37/1.12      & {\bf 0.24/0.04}   & \bf{1.18/0.49}  \\

    \hline
  \end{tabular}
}
\end{center}

\label{table:table1_res}
\end{table}

 \section{Results and Discussions}
 \label{sec:res_diss}
 This section presents experimental results on TD-SV. 
 
\subsection{Results under the GMM-UBM framework}

Fig. \ref{fig:sv_eer_alpha} shows TD-SV results of MFCC, Spk-BN and APC-BN features and their score-level fusion for different VTL factors ($\alpha$)  on the RedDots database. 
We observe that APC-BN yields lowest EER values for Target-wrong and Imposter-wrong over all values of $\alpha$. However, it fails in Imposter-correct as compared to MFCC and Spk-BN. This can be explain by the fact that the APC objective is to predict the content of future frames and thus the APC-BN feature is sensitive to the linguistic content. On average, APC-BN significantly outperforms other features. Fusion of three feature systems further reduces the EERs.  

Table \ref{table:table1_res} presents the results of the proposed VTL perturbation method and its combination with APC-BN and Spk-BN features. It is seen that the proposed method significantly reduces the average EERs for all features. Fusion of the feature-dependent systems further improves the performance; the EER reduces from $2.52\%$ of the baseline MFCC to $1.18\%$ of the fused system.  Similar observation is seen in terms of minDCF. Note that the average EER for MFCC is 2.52\% as compared to 3.19\% in \cite{DBLP:journals/taslp/SarkarTTSG19}, and the improvement is due to the use of rVAD \cite{TanSD20} in this work instead of an energy based one.

\begin{table}[!b]
\caption{\it TD-SV performance of the proposed VTL perturbation methdo and its combination with APC-BN and Spk-BN features on the RedDots database in terms of [\%EER/(MinDCF$\times$ 100)]) with i-vector.}
\setlength\tabcolsep{1.70pt}
\begin{center}
\scriptsize{
\begin{tabular}{|llll|l|}\cline{1-5}
System & \multicolumn{3}{c|}{Non-target type} & Avg. \\
       & Target-      & Imposter & Imposter- & EER/\\ 
      &  wrong        & correct  &  wrong    & MinDCF\\ \hline
{\bf Baseline:} & & & & \\
MFCC ($\alpha=1.0$) &6.07/2.60  & 4.07/1.81 & 1.17/0.46   & 3.77/1.62\\ 
Spk-BN ($\alpha=1.0$) &  6.39/2.78& 4.87/1.98& 1.60/0.63 & 4.29/1.80  \\ 
APC-BN ($\alpha$=$1.0$)         &2.07/0.65	&7.97/3.22 &0.77/0.28      & 3.60/1.38\\
         &    & & & \\       
{\bf Proposed:} & & & & \\ 
MFCC (all $\alpha$)          &4.88/2.10 &2.80/1.23 &0.67/0.25      & 2.78/1.19 \\  
Spk-BN (all $\alpha$)  & 4.07/1.85  & 2.96/1.28 &0.60/0.23 & 2.54/1.12\\ 
APC-BN (all $\alpha$) &  1.10/0.33 & 6.07/2.18  & 0.46/0.09 &  2.54/0.87  \\
& & & & \\
{\bf Score fusion:} & & & & \\ 
MFCC + Spk-BN ($\alpha$=$1.0$) & 3.26/1.22       &  {\bf 2.51/1.09}    & 0.77/0.19 & 2.18/0.84\\
MFCC + APC-BN ($\alpha$=$1.0$) &  1.69/0.58    &   5.27/2.12   &  0.49/0.13  & 2.48/0.94\\
MFCC + [Spk + APC]-BN ($\alpha$=$1.0$) &  1.56/0.58    & 4.10/1.68     &  0.43/0.11  & 2.03/0.79\\
 & & & & \\
MFCC + Spk-BN (all $\alpha$)  & 4.27/1.88 &   2.71/1.16 &0.57/0.21  & 2.51/1.08  \\
MFCC + APC-BN (all $\alpha$) & {\bf 0.98/0.35}   & 4.03/1.47  & {\bf 0.33/0.06} & 1.78/0.63 \\
MFCC+ [Spk + APC]-BN (all $\alpha$) &  1.07/0.36  &  3.28/1.23 & {\bf 0.33/0.06} & {\bf 1.56/0.55} \\
  \hline
  \end{tabular}
}
  \end{center}
\label{table:table2_res}
\end{table}

\subsection{Results under the i-vector framework}

Table \ref{table:table2_res} presents the corresponding results using the i-vector technique, and we can see similar observations to those of GMM-UBM. 

\section{Conclusion}
\label{sec:con}
In this letter, we proposed a VTL perturbation method for TD-SV and experimentally demonstrated its effectiveness. We further explored the performance of APC-BN for TD-SV in comparison to the cepstral feature and the speaker-discriminant BN feature, and APC-BN showed superior performance. The VTL perturbation method was then applied to BN features, which improved the performance further. In the end, fusion of three feature-dependent systems performs the best. 

\newpage

\bibliographystyle{IEEEbib}
 \bibliography{strings,References}

\end{document}